\documentclass{svmult}
\usepackage{amsmath,bm,upgreek,cite}

\usepackage{mathptmx}       
\usepackage{makeidx}         
\usepackage{graphicx}        
\DeclareGraphicsExtensions{.pdf}
\usepackage{multicol}        
\usepackage[bottom]{footmisc}

\makeindex             

\spnewtheorem*{anomaly}{Anomaly}{\bf}{\it}
\spnewtheorem*{goal}{Goal}{\bf}{\it}
\spnewtheorem{principle}{Principle}{\bf}{\it}
\newcommand{\beq}{\begin{equation}}
\newcommand{\eeq}{\end{equation}}
\newcommand{\beqa}{\begin{eqnarray}}
\newcommand{\eeqa}{\end{eqnarray}}
\newcommand{\beqan}{\begin{eqnarray*}}
\newcommand{\eeqan}{\end{eqnarray*}}

\newcommand{\ket}[1]{|{#1}\rangle}

\newcommand{\eq}[1]{Eq.~(\ref{#1})}

\begin{document}

\title*{An anomaly in space and time and the origin of dynamics}
\author{Joan A. Vaccaro}
\institute{Joan A. Vaccaro \at Centre for Quantum Dynamics, School of Natural Sciences, Griffith University, Brisbane, Australia, \email{J.A.Vaccaro@griffith.edu.au}}

\maketitle

\abstract{
The Hamiltonian defines the dynamical properties of the universe.  Evidence from particle physics shows that there is a different version of the Hamiltonian for each direction of time.  As there is no physical basis for the universe to be asymmetric in time, both versions must operate equally. However, conventional physical theories accommodate only one version of the Hamiltonian and one direction of time. This represents an unexplained anomaly in conventional physics and calls for a reworking of the concepts of time and space. Here I explain how the anomaly can be resolved by allowing dynamics to emerge phenomenologically.  The resolution offers a picture of time and space that lies below our everyday experience, and one in which their differences are epiphenomenal rather than elemental.
}

\section{Introduction}
One of the earliest attempts to describe the nature of time and space comes from Parmenides ($\sim$ 500 BCE) \cite{Plato}. He and his pupil Zeno argued for monism---that there was only a single reality---and so to them time was a complete whole without division. They argued that this gives less absurdities than the opposing pluralistic view where multiple realities catered for different modes of being.  Zeno's well-known paradoxes were attempts to illustrate the absurdities that would follow from pluralism.  However, a new way of looking at nature, based on empirical observations and mathematical calculations, emerged in the European Renaissance period. The perceived difficulties associated with Zeno's paradoxes were largely swept aside with the development of calculus. Building on the work of Copernicus and Galileo, Newton proposed that an absolute time flows uniformly throughout an absolute space \cite{Novikov}.  Newton's framework represents a kind of pluralism where each moment in time represents a separate reality.  Then, about a century ago, James \cite{James} and McTaggart \cite{McTaggart} resurrected a monist view of time in the form of the block universe, where time is seen to be one structure without a present, past or future. The block universe represents, for the most part, the orthodoxy among physicists in modern times \cite{Zeh,Price}. Nevertheless, a new kind of pluralism will reemerge later in this chapter.

The impetus for abandoning Newton's framework of space and time in physics came from its failure to account for the propagation of light in the Michelson-Morley experiments of 1887. This anomaly led Einstein in 1905 to propose a new framework for space and time in his special theory of relativity. How we think of time and space today in terms of a background geometry is moulded by Einstein's relativistic \emph{spacetime}---an amalgamation of space and time into a single entity.  An interval of either time or space for one reference frame can be an interval that extends over both time and space in another reference frame.\footnote{Appendix 1 discusses this in more detail.}
 In this sense, one can say that space and time appear in special relativity on the same footing.

Yet time and space are quite different in other respects. For example, matter can be localised in a region of space but not in an interval of time. That is, a lump of matter---such as an atom, a coffee cup or even a galaxy---can exist in one region of space and no other, but conservation of mass\footnote{The terms mass and matter here can be taken to mean relativistic energy.} forbids matter from existing at one time interval and no other.
To exist only at one time interval, for one second after midday say, would mean the matter not existing before midday, existing only during the second after midday and vanishing at the end of the second. We avoid this drastic violation of mass conservation in conventional physics by insisting that matter follows an \emph{equation of motion} that translates it over all times.
The upshot is that matter is presumed to undergo continuous translation over time (as time evolution) but there is no corresponding presumption about the matter undergoing translations over space.

There is more to this---the presumed continuous translation over time occurs in a preferred direction and the direction is described by various arrows of time.  The first to be named formally is the thermodynamic arrow \cite{Eddington} which points in the direction of increasing entropy.  Other arrows include the cosmological arrow, which points away from the big bang, and the radiation arrow, which points in the direction of emission of waves \cite{Price}.
In contrast, space is isotropic.

There is another, quite subtle, difference between time and space that has largely escaped attention until recently: translations in time and space have very different discrete symmetry properties \cite{FPhys,FPhys2,PRSA}.  The discrete symmetries here represent an invariance to the operations of charge conjugation (C), parity inversion (P) and time reversal (T). Although nature respects these symmetries in most situations, exceptions have been discovered in the last 60 years.  The exceptions are observed as violations of particular combinations of the C, P and T symmetries in certain particle decays \cite{parity,CP,T1,T2,T3,T4}. The violations are independent of position in space, and so they occur over \emph{translations in time} (i.e. as a decay) and \emph{not translations in space}.

The fact that time and space have these differences does not, in itself, constitute a problem.  On the surface, the differences don't appear to be pointing to a glaring anomaly that requires a reworking of the foundations of physics like the results of the Michelson-Morley experiment did. Yet there is an anomaly, one that has been around for so long that it risks being overlooked because of its familiarity.  It is to do with the fact that there is no cause for the block universe to be anything other than symmetrical in time. In other words, there is no physical basis for one direction of time to be singled out \cite{Price}.  This invites the question, so where is the other direction of time?  It may be tempting to speculate that another part of the time axis may carry arrows pointing in the opposite direction.  But this will not do, given the impact the discoveries of the violation of the discrete symmetries have for the Hamiltonian. The Hamiltonian is a mathematical object that defines the dynamics. The violation of time reversal symmetry, called T violation for short, implies that there is a different version of the Hamiltonian for each direction of time, yet we observe only one version in our universe and, not surprisingly, only the observed version of the Hamiltonian appears in conventional theories of physics. Where is the other direction of time and its concomitant version of the Hamiltonian? The fact that there is no answer in conventional physics constitutes a basic anomaly which calls for a fundamental shift in our thinking about time and space.

The purpose of this chapter is to expose the anomaly and then review my recent proposal \cite{PRSA} to resolve it through restructuring the way time and space appear in physical theory. The anomaly is articulated more precisely in \S2 and then \S 3 prepares for the required restructuring in terms of a goal and three basic principles.  Following that, the principles are applied to non-relativistic quantum mechanics in \S 4 and the chapter ends with a discussion in \S 5.  Additional background material and specific details are left to the appendices: special relativity in Appendix 1, generators and translations in time and space in Appendix 2, and quantum virtual paths in Appendix 3.  Full details of my proposed resolution can be found in Ref.~\cite{PRSA}. 

\section{An anomaly: missing direction of time and its Hamiltonian}

To expose the anomaly we must first lay to rest a common misconception that the arrows of time are melded in some way into the concept of time itself.
In particular, if the only thing that distinguishes the two directions of time is an increase of entropy in one direction, then perhaps one could be forgiven for succumbing to a conceptual shorthand and regarding the entropy increase as somehow causing the direction of time.
But, in truth, the arrows are only \emph{evidence} that time has a direction and there is simply no basis for claiming them as the \emph{cause} of that direction.
An analogy will help make the distinction between evidence and cause clearer.  Imagine that the leaves falling from a tree are blown by a steady wind to land preferentially on the downwind side of the tree.
The pattern of leaves on the ground would then provide \emph{evidence} of the direction that the wind is blowing, but there would be no basis for claiming that the leaves \emph{cause} the wind to have any particular direction.
The same situation occurs with the direction of time: the arrows are patterns that provide evidence of the direction of the translations over time, but those patterns do not cause the translations themselves nor do they cause the translations to be in a particular direction.

Having laid bare the evidential nature of the arrows, we now examine the thermodynamic arrow in particular.  This arrow, like all the arrows, is phenomenological in origin.  It arises because thermodynamics was developed to be in accord with nature and thus it was intentionally structured to have an increasing entropy in the direction of time we refer to as ``forwards'' or the ``future''.  However, as Loschmidt pointed out long ago, thermodynamics is consistent with time-symmetric physical laws, such as Newton's laws of motion, and so any prediction of an increase in entropy in one direction of time is, necessarily, a prediction of an increase in the opposite direction of time.  To ignore this and claim that the thermodynamic arrow, or any of the arrows, explains the direction of time, is to commit what Price calls a double standard fallacy \cite{Price}.  Avoiding the fallacy leaves us with the problem of a missing direction of time.

Its resolution calls for a time-symmetric model of nature that accounts for both directions of time---a model in which there are reasons for arrows to point in both directions.  There have been admirable attempts along these lines by Carroll, Barbour and their co-workers \cite{Carroll,Barbour},  but there is something fundamental missing from their analyses because they only consider time-symmetric physical laws. The only fundamental law that is not time symmetric is usually dismissed as having little to do with large-scale effects \cite{Novikov,Price,Zeh,Carroll-book}.  It is associated with the weak interaction, and its time asymmetry is observed as T violation in the decay of the K and B mesons \cite{T1,T2,T3,T4}.  However, despite being previously overlooked, I have shown that T violation is capable of producing large-scale physical effects \cite{FPhys,FPhys2,PRSA}.  Moreover, the experimentally observed T violation implies that the universe is described by two versions of the Hamiltonian, one for each direction of time.  The double-headed arrows of Carroll, Barbour and co-workers do not account for this crucial fact.

The problem, then, is not only that there is a missing direction of time, but that the associated version of the Hamiltonian is missing along with it.  The anomaly is the rather glaring absence of both directions of time and both versions of the Hamiltonian in conventional physical theories; it can be stated formally as follows.
\begin{anomaly}
There is no basis for nature to be asymmetric in time.  Experiments in particle physics indicate that there are two versions of the Hamiltonian, one for each direction of time.  A time-symmetric theory of nature must give an equal account of both directions of time and both versions of the Hamiltonian.  Conventional theories fail in this regard because they can accommodate only one version of the Hamiltonian and one direction of time.
\end{anomaly}
The anomaly calls for a restructuring of the concepts of time and space in physics.

\section{The goal and basic principles}

The goal of the restructuring might appear to be to simply find a time-symmetric description that includes both versions of the Hamiltonians.  However, aiming the goal directly at the anomaly like this misses an opportunity for rebuilding from a deeper level. For example, if Einstein had been satisfied with a description of the propagation of light that was consistent with the Michelson-Morley experiment, he may have settled on some aether-dragging model. Instead, his search for an indirect, but deeper, solution led to his special theory of relativity, a natural consequence of which was the resolution of the light-propagation anomaly.  In the same way, we need to take a step back from the anomaly itself.  We have seen that the differences between time and space are related by the fact that they involve translations:  conservation laws and the equation of motion represent translations over time, the direction of time describes an asymmetry in translations over time, and the violation of the discrete symmetries is observed for translations over time.  Our understanding of the relationship between time and space would be advanced significantly if all differences could be shown to have a common origin.  The least understood among the differences is the C, P and T symmetry violations.  Although the violations are generally considered to represent profound properties of nature, they don't play any significant role in conventional physics.  Indeed, they stand out as having been overlooked.  To address this situation, we undertake the more ambitious goal as follows:
\begin{goal}
To treat time and space on an equal footing at a fundamental level, and to allow their familiar differences to emerge phenomenologically from the discrete symmetry violations.
\end{goal}
If the violations deliver the differences between space and time then we will have found a theory that incorporates both versions of the Hamiltonian in a way that gives rise to the familiar direction of time. The anomaly would then be resolved as a natural consequence of the goal.

Having settled on the goal, we now turn to the basic principles needed to achieve it.
When the C, P and T symmetries are obeyed we want matter to be localisable both in time and space.  This will require a formalism in which conservation laws do not apply and an equation of motion is not defined---this marks a serious departure from conventional physics.  When the violation of the symmetries are introduced into the formalism, an effective equation of motion and conservation laws need to appear phenomenologically as a consequence---only then will it be in agreement with conventional physics.  The symmetry violations clearly need to play a significant role in the formalism.  The violations manifest as changes due to the C, P and T operations,\footnote{If the C, P and T operations do not change the system then the symmetries are obeyed.  Violations represent the converse situation where changes result from the operations.} and so their impact would tend to be greater in a formalism in which the operations are more numerous.  The P and T operations, in particular, are associated with reversing directions in space and time, respectively.  It is clear from this that we need a formalism comprising paths in time and space which suffer innumerably-many reversals.  A stochastic Wiener process involves paths of this kind in space.  Feynman's path integral method \cite{Feynman} also involves similar kinds of paths over configuration space.

The important point about Feynman's method is that it underpins analytical mechanics in the limit that Planck's constant, $\hbar$, tends to zero.  Indeed, his method shows that Hamilton's principle of least action arises as a consequence of destructive interference over all possible paths in configuration space between the initial and final points.  But it stops short of considering paths that zigzag over time of the kind we need to consider here and, as a consequence, it stops short of considering the impact of the C, P and T symmetry violations that are the focus here.  Nonetheless, it does demonstrate the importance of quantum path integrals for describing the universe on a large scale.

Although the paths need to comprise innumerably-many reversals, there are reasons to believe that there are physical limitations to the resolution of intervals in space and time \cite{Planck}.  For example, the position of an object can be determined by observing the photons it scatters, but the accuracy of the result cannot be better than the Planck length $L_P=1.6\times 10^{-35}$~m \cite{Mead}.   Correspondingly, the timing of the scattering events cannot be determined any better than the Planck time $L_T=L_P/c=5.4\times 10^{-44}$~s where $c$ is the speed of light.  We will assume that fundamental resolution limits of this kind exist without specifying their value. It would be physically impossible to resolve the structure of paths with step sizes smaller than the resolution limit, and so we need to treat such paths as having equal physical status.

With these ideas in mind we formulate three principles on which to base the development of the new formalism:
\begin{principle}  \label{p:qvp}
A quantum state is represented as a superposition of paths, each containing many reversals. We call these ``quantum virtual paths''.
\end{principle}
\begin{principle}  \label{p:resolution}
There is a lower limit to the resolution of intervals in space and time.  Quantum virtual paths with step sizes smaller that this limit have an equal physical status.
\end{principle}
\begin{principle}  \label{p:phenomenological}
States have the same construction in both time and space. Any differences between space and time, such as dynamics and conservation laws, emerge phenomenologically as a result of the violation of discrete symmetries C, P and T.
\end{principle}

\section{Applying the principles}

\begin{figure}[t]
\sidecaption
  \includegraphics[width=75mm]{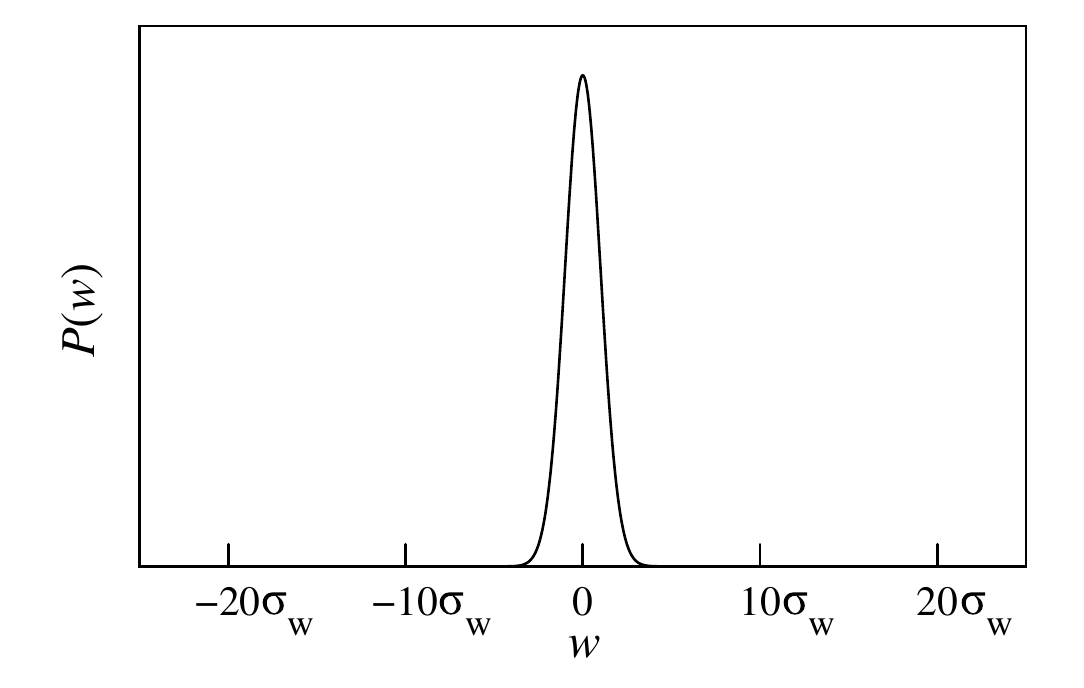}
  \caption{Bell-shaped probability distribution $P(w)$ representing an object localised in the vicinity of the origin of the $w$ coordinate. The standard deviation of the distribution is $\sigma_{\rm w}$.}\label{fig:gaussian}
\end{figure}

We shall apply the three basic principles to represent the quantum state of an object.\footnote{We only use the static representation of a state from non-relativistic quantum mechanics.  We do not apply an equation of motion nor do we impose conservation laws.} The object represents the only matter in space and time and it could be an atom, planet or galaxy. Its details are not important.  We will refer to it as the ``galaxy'' in the following.  The first task is to develop the formalism in general terms without referring specifically to time or space. For that let $w$ be a generic coordinate which will later be set to be either time or space. We want the galaxy to be localised with respect to $w$ such that the spread in $w$ is finite.   The most general probability distribution with a finite spread has a bell-shape like $P(w)$ illustrated in Fig. \ref{fig:gaussian}.

\subsection{Application of principle \ref{p:qvp}}

An equivalent representation is given by imagining that the galaxy takes a path that starts at the origin $w=0$ and randomly steps back and forth along the $w$ coordinate a number of times. Let there be $N$ steps in the path and let the magnitude of each step be $\delta w$.
For the final location of the galaxy to any value of $w$, the step size $\delta w$ needs to be infinitesimally small and $N$ needs to be correspondingly large.  By setting
\begin{align}   \label{eq:delta w}
     \delta w = \frac{\sqrt{2}\sigma_{\rm w}}{\sqrt{N}}
\end{align}
and choosing a suitably-large value of $N$ we can make the step size, $\delta w$, as small as we like, and the maximum length of any path, $N\delta w$, correspondingly as large as we like, while keeping the standard deviation in the possible final locations fixed at $\sigma_{\rm w}$.
It needs to be emphasised that even though temporal references such as ``starts'', ``steps'' and ``final'' are used here, the paths do not represent actual movement over a time interval.  Rather they represent the galaxy executing a sequence of \emph{virtual displacements} along $w$ without any reference to time at all.
That is, the galaxy is considered to be simply displaced from $w=0$ to the point represented by the end of the random path.
Virtual displacements arise in analytical mechanics when discussing constraints on motion \cite{Goldstein}; here the accumulation of many random virtual displacements give the possible values of $w$.

\begin{figure}[t]
  \sidecaption
  \includegraphics[width=75mm]{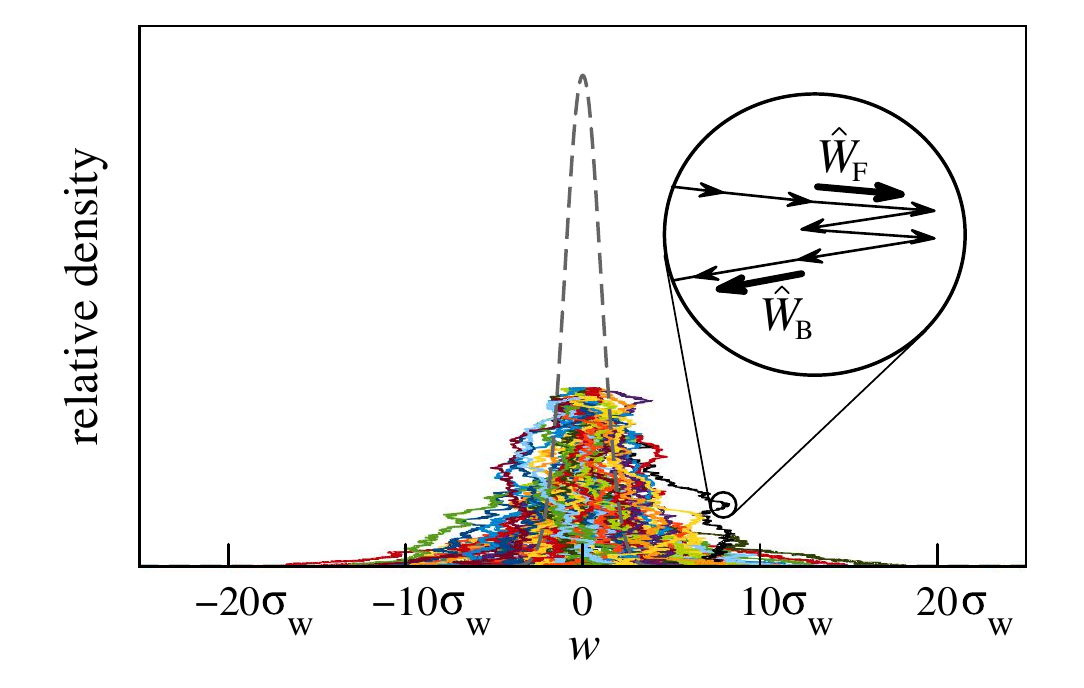}
  \caption{Conceptual sketch of a quantum virtual path. Each curve represents a random path of $N$ steps back and forth along the $w$ coordinate starting at $w=0$ and ending at a random value of $w$. The curves are displaced vertically to represent the relative density of paths. The inset illustrates the actions of the generators, $\hat W_F$ and $\hat W_B$, of translations in the $+w$ and $-w$ directions, respectively.  }\label{fig:qvp - w - symmetry case}
\end{figure}

For the location of the galaxy to be described by the smooth bell-shaped distribution $P(w)$ we need not just one path and its end point, but infinitely many. We don't know which end point describes the location of the galaxy and so we have to allow for the possibility that it could be the end point of any one of many paths.  Technically, this means we represent the location of the galaxy by a \emph{superposition} of the end points of all the paths.
The superposition is called a ``quantum virtual path'', where quantum refers to the fact that it is a quantum superposition \cite{PRSA}.

One can imagine a quantum virtual path for a specific value of $N$, say $N=600$, as the sum of the end points of all the paths illustrated in Fig. \ref{fig:qvp - w - symmetry case}.  The step size $\delta w$ for each zigzag path in the figure is given by \eq{eq:delta w} for some fixed value of the standard deviation $\sigma_{\rm w}$.  Another quantum virtual path can be constructed for $N=601$ in a similar way for a correspondingly smaller step size $\delta w$.  Imagine that this has been done for every positive integer value of $N$.
As $N$ increases in this imagined process, the step size $\delta w$ reduces and the quantum virtual path represents an ever finer description of the state of the galaxy, eventually tending to the bell-shaped dashed curve shown in the figure.  Each quantum virtual path so constructed represents a possible state of the galaxy in terms of its location along the $w$ coordinate.

Each step of $\delta w$ is produced using a particular operation called a ``generator'' of the translation.  In particular, $\hat W_F$ is the generator for translations that increase the value of $w$ and $\hat W_B$ is the generator for ones that decrease its value, as illustrated in the inset of Fig.~\ref{fig:qvp - w - symmetry case}.  If the generators are invariant to reversals of direction then they are equivalent, i.e. $\hat W_F=\hat W_B$.  More will be said about this later.
A technical review of generators and translations is given in Appendix 2 and a brief discussion of how a quantum virtual path is related to the bell-shaped distribution $P(w)$ can be found in Appendix 3.

\subsection{Application of principle \ref{p:resolution}}

As the value of $N$ increases, the step size $\delta w$ from \eq{eq:delta w} becomes smaller.  At some point $\delta w$ will be smaller than the resolution limit $\delta w_{\rm min}$ for the $w$ coordinate.
All quantum virtual paths with a step size $\delta w$ smaller than $\delta w_{\rm min}$ will give descriptions of equal status according to Principle \ref{p:resolution}.
For convenience, we shall collect the equivalent quantum virtual paths in a set called $\mathbf{G}$.  Each quantum virtual path in this set equally represents the state of the galaxy in terms of its location along the $w$ coordinate.  There are an infinite number of such quantum virtual paths in the set $\mathbf{G}$.

\subsection{Application of principle \ref{p:phenomenological}}

We now discuss space and time explicitly. First consider the spatial case which, for brevity, we limit to just the $x$ dimension.  In this case the generic coordinate $w$ is replaced with $x$ and the generator of translations is replaced with $\hat p_x$, the component of momentum along the $x$ axis.  There is only one generator for translations in both directions of the $x$ axis and so $\hat W_F=\hat W_B=\hat p_x$ here.  Further technical details are given in Appendix 2.   Fig.~\ref{fig:qvp - w - symmetry case} with $w$ replaced by $x$ illustrates a quantum virtual path over the $x$ axis.  Collecting the quantum virtual paths with a step size smaller than some minimum resolution limit yields the set of states of equal status which we will call ${\bm{\upPsi}}$.   All the quantum virtual paths in $\bm{\upPsi}$ are physically indistinguishable from the bell-shaped distribution $P(x)$ represented in Fig.~\ref{fig:gaussian} with $w$ replaced with $x$.

Next, we repeat the same exercise for time.  In this case the coordinate is $w=t$ and, in general, there are two generators of translations given by the two versions of the Hamiltonian, i.e. $\hat W_F=\hat H_F$ and $\hat W_B=\hat H_B$ corresponding to the ``forwards'' and ``backwards'' directions of time, respectively. Technical details regarding these generators are given in Appendix 2.  As with the spatial case, Fig.~\ref{fig:qvp - w - symmetry case} with $w$ replaced by $t$ illustrates a quantum virtual path over the $t$ axis, and
collecting the quantum virtual paths which have a step size smaller than some minimum resolution limit yields the set of states of equal status which we will call ${\bm{\upUpsilon}}$.

In a universe where the T symmetry holds, there is only one version of the Hamiltonian and so $\hat H_B=\hat H_F=\hat H$. In this case the galaxy is localised in time within a duration of the order of $\sigma_{\rm t}$ of the origin and all the states in $\bm{\upUpsilon}$ are physically indistinguishable from the bell-shaped distribution $P(t)$ represented in Fig.~\ref{fig:gaussian} with $w$ replaced with $t$.
The galaxy only exists in time for a relatively short duration at the origin $t=0$ and does not exist before or after this time.  It can be imagined to come into existance momentarily and then promptly vanish.
Clearly, in this case, the galaxy has the same representation in time as in space---it is localised in both---and the formalism places time and space on the same footing in this respect.  This is far removed from conventional quantum mechanics as there is no equation of motion and the mass of the galaxy is not conserved.

\begin{figure}[t]
  \sidecaption
  \includegraphics[width=75mm]{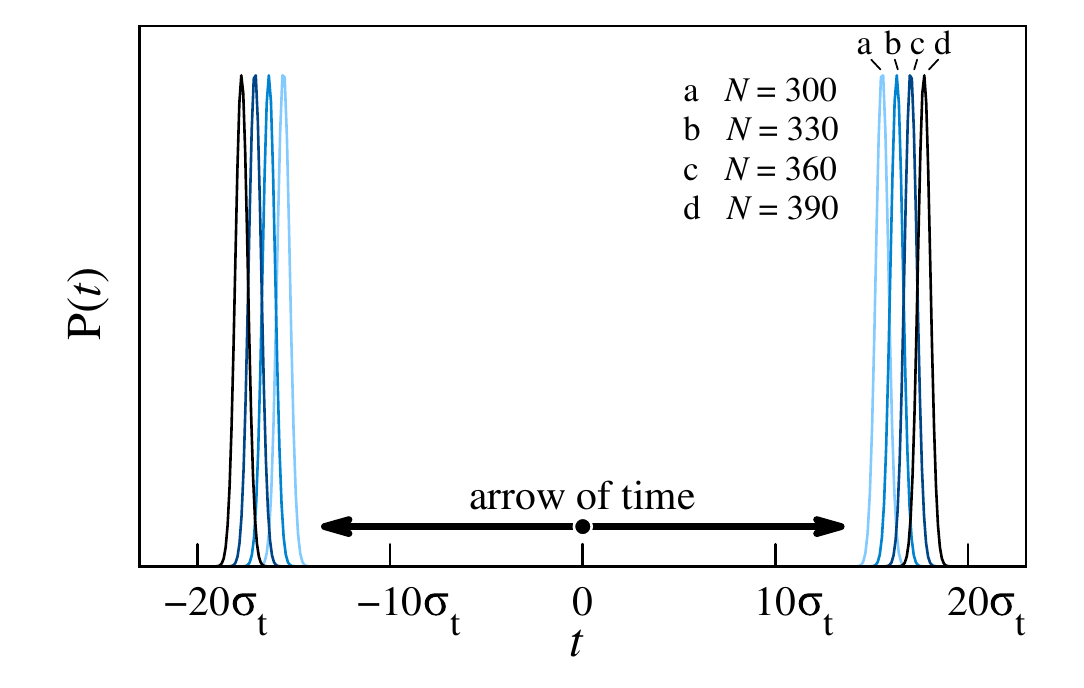}
  \caption{The probability distribution $P(t)$ for various values of $N$ in the case of T violation. Each curve has two bell-shaped peaks which move further apart as $N$ increases.  }\label{fig:qvp - t - violation}
\end{figure}

The converse case, where T symmetry is violated, is defined by $\hat H_B\ne\hat H_F$.   The key point here is that multiple paths that zigzag in different ways from the origin to the same end point can interfere.  The interference can be compared to the way waves travelling on the surface of water behave; if the trough of one wave occurs at the same point as the crest of another, the two waves will tend to cancel each other in a process called destructive interference, whereas if two troughs or two crests meet they tend to reinforce each other as deeper troughs or higher crests, respectively, in a process called constructive interference.  In a similar way, multiple paths that end at the same point on the time axis interfere either destructively or constructively.  The result is that instead of the probability distribution having a maximum at the origin, like the bell-shaped curve in Fig.~\ref{fig:gaussian}, destructive interference reduces the probability to zero in this region.  This is compensated by constructive interference that yields two symmetrically positioned bell-shaped peaks further from the origin as illustrated in Fig.~\ref{fig:qvp - t - violation}.
 In other words, each quantum virtual path is now composed of two bell-shaped peaks that represent the galaxy existing at two different times, $+t$ and $-t$, say.  This situation is like Schr\"{o}dinger's cat that exists in a superposition of being both dead and alive simultaneously, except that here the galaxy is at two different times.  As the value of $N$ increases, the two peaks become further separated as shown in Fig.~\ref{fig:qvp - t - violation}, and the galaxy shifts in time accordingly.  Each quantum virtual path represents one of the states in the set  ${\bm{\upUpsilon}}$, and according to Principle \ref{p:resolution}, has equal physical status.  In terms of Fig.~\ref{fig:qvp - t - violation}, this means that each double-peaked curve equally represent the position of the galaxy in time.

The presence of T violation clearly has a dramatic affect on the temporal description of the galaxy.  For example, consider the question, where in time is the galaxy likely to be found?  Without T violation, the unequivocal answer is only near the origin in accordance with Fig.~\ref{fig:gaussian}, whereas with T violation, the answer implied by Fig.~\ref{fig:qvp - t - violation} would be at \emph{any} time $t$.

\subsection{The origin of dynamics}

We now focus on the T violation case.  According to Principle \ref{p:resolution}, all states in the set ${\bm{\upUpsilon}}$ have an equal status in representing the galaxy in time.  For any given value of time, $t$, there is a corresponding state in ${\bm{\upUpsilon}}$ that represents the galaxy being in a superposition of the times $+t$ and $-t$.\footnote{In principle, the time $t$ could be chosen to be the current age of the universe, $13.8$ billion years. There is a state in ${\bm{\upUpsilon}}$ that represents the galaxy being in a superposition of the times 13.8 billion years and $-13.8$ billion years.}   This implies that the galaxy exists at any time we wish to consider, and so its \emph{mass is conserved}. This conservation law has not been imposed on the formalism, as it would need to be in conventional theories, but rather it is phenomenology arising from T violation.

The corresponding equation of motion is found as follows. The two peaks $+t$ and $-t$ in each curve in Fig.~\ref{fig:qvp - t - violation} represent time-reversed versions of the galaxy.
An observer in the galaxy would not be able to distinguish between them and so we need only consider one, at $+t$ say.  If the observer makes observations with a resolution in time that is broader than the width of the peak, the peak will appear to be instantaneous and a set of them will appear to form a continuous sequence.  Under these circumstances, the observer would find evidence of an \emph{equation of motion} that is consistent with the Schr\"{o}dinger equation of conventional quantum mechanics. This equation has not been imposed on the formalism but rather it arises as phenomenology associated with T violation. This suggests that the origin of dynamics lies in T violation.

The remaining distinctive feature of time to consider is its direction, and the states in ${\bm{\upUpsilon}}$  have a time ordering in the following sense.  According to the meaning of time evolution defined in Appendix 2, the peak labelled ``d'' Fig.~\ref{fig:qvp - t - violation} represents a state that has \emph{evolved in time} from the state represented by the peak labelled ``c'', and that state has evolved from the state represented by ``b'', which has evolved from the state represented by ``a'', but the converse is not true. This means that there is an arrow of time pointing in the direction of $+t$.  The same argument applies to the time reversed states in regards to the $-t$ direction and so the \emph{arrow is double headed}, like those of Carroll and Barbour and co-workers \cite{Carroll,Barbour}.  The important point here is that both versions of the Hamiltonian, $\hat H_F$ and $\hat H_B$, are included in the formalism.

We have now achieved our goal: we treated time and space on an equal footing and found their familiar differences to emerge phenomenologically from T violation.

\section{Discussion}

We began by identifying a fundamental anomaly in physics, viz. conventional theories fail to give a time symmetric description that accounts equally for both versions of the Hamiltonian and both directions of time.  We have presented a new formalism for quantum mechanics that resolves this anomaly.  The new formalism is based on three principles that allow quantum states in time and space to be treated on an equal footing in terms of quantum virtual paths.  The distinctive features associated with time, i.e. conservation laws, equation of motion and the direction of time, are not imposed on the formalism but rather emerge phenomenologically as a result of T violation. These key differences between time and space follow from the fact that the generators of translations in space and time, the momentum operator and the Hamiltonian, respectively, have different symmetry properties: the momentum operator is invariant to the C, P and T symmetry operations whereas the Hamiltonian is not. Accounting for these differences gives the \emph{origin of dynamics}.

The new formalism also refines the meaning of time.  In conventional theories, the word ``time'' refers to both a coordinate of a space-time \emph{background} as well as the parameter describing \emph{dynamical evolution}. Both concepts are firmly entwined by conservation laws.  For example, the conservation of mass implies that a massive object will persist over all times and, accordingly, it is represented on a space-time background as existing at each time. The dynamical evolution of the object becomes the path of the object on the space-time background. Here, however, the two concepts of time as a background coordinate and as a dynamical parameter are distinct.  Time and space have an equal footing as a background on which quantum states are represented.  The states, as quantum virtual paths, represent objects that are localised in time and space: each state in the sets ${\bm{\upPsi}}$ and ${\bm{\upUpsilon}}$ represents a relatively-narrow bell-shaped distribution or a sum of two relatively-narrow bell-shaped distributions.  In particular, mass is not conserved and there is no equation of motion for any \emph{individual state} in ${\bm{\upUpsilon}}$ (as illustrated by Fig.~\ref{fig:qvp - t - violation})---time appears only as a background coordinate. In contrast, mass conservation, the equation of motion and the direction of time, are properties of the\emph{ whole set} ${\bm{\upUpsilon}}$ where time appears as a dynamical parameter.  In other words, time as a background coordinate and as a dynamical parameter apply to distinct constructs in the formalism.

It might appear unusual that a quantum formalism is being proposed to explain large scale structure of nature given that quantum effects are typically seen only in relatively small systems under controlled conditions.  However, Feynman's path integral method has already demonstrated how quantum phenomena underpins Hamilton's least action principle in analytical mechanics \cite{Feynman} and thus large scale structure. In this regard, the new formalism should be considered as an extension of Feynman's method to encompass paths over time and the C, P and T symmetry violations and, thus, to apply to nature on a large scale as well.

Finally, the set of states ${\bm{\upUpsilon}}$ for T violation represents the galaxy at an infinite sequence of times. Each state in ${\bm{\upUpsilon}}$ may be viewed as representing a different reality.  In this sense, the formalism resurrects a kind of pluralism.  The monism-pluralism cycle for time turns once more.

\section*{Appendix 1}
\addcontentsline{toc}{section}{Appendix 1}
We briefly review here the Lorentz transformation in special relativity. A point in spacetime is referred to as an event; it is specified by four coordinates $x$, $y$, $z$, $t$ with respect to a reference frame.
The Lorentz transformation gives the relationship between the coordinates of two different inertial reference frames. In particular, consider two events that occur a distance of $\Delta x$ apart along the $x$ axis and separated by a duration of $\Delta t$ in time in the  $x$, $y$, $z$, $t$ reference frame.
In the $x'$, $y'$, $z'$, $t'$ reference frame that is moving a constant speed $v$ along the $x$ axis of the first, the distance and duration along the $x'$ and $t'$ axes between the events are given by
\begin{align*}
       \Delta x' &=\gamma(\Delta x-v\Delta t)\\
        \Delta t' & =\gamma(\Delta t-v\Delta x/c^2)\ ,
\end{align*}
respectively, where $\gamma =\sqrt{1-v^2/c^2}$ and $c$ is the speed of light.
The important point here is that what is considered to be solely a spatial interval, $\Delta x$, in one reference frame becomes part of a temporal interval $\Delta t'$ as well as being part of a spatial interval $\Delta x'$ in the other reference frame.  That is, space and time are interchangeable.

\section*{Appendix 2}
\addcontentsline{toc}{section}{Appendix 2}

Here, we briefly review translations and their generators. Recall that the Taylor expansion of a function $f(x)$,
\[
     f(x+a) = f(x) + a\frac{\D}{\D x}f(x) + \frac{a^2}{2!} \frac{\D^2}{\D x^2}f(x) + \frac{a^3}{3!} \frac{\D^3}{\D x^3}f(x) + \ldots\ ,
\]
can be written compactly in exponential form as
\begin{align*}
        f(x+a) = \E^{-\I a(\I\frac{\D}{\D x})}f(x)\ .
\end{align*}
When written in this form the differential operator $\I\frac{\D}{\D x}$ is said to be the generator of translations in $x$.  The generator of spatial translations along the $x$ axis is $\hat p_x$, the operator representing the $x$ component of momentum.  We need only consider one dimension of space for our purposes here.  Thus we write
\begin{align}    \label{eq:x - translate by a}
        \ket{x+a}_{\rm x}=\E^{-\I a\hat p_x}\ket{x}_{\rm x}
\end{align}
where $\ket{x}_{\rm x}$ represents a state vector for position $x$ and, for convenience, we assume units in which $\hbar=1$. Similarly, the generator of translations in time $t$ is the Hamiltonian operator $\hat H$ and so
\begin{align}    \label{eq:t - translate by a}
        \ket{\psi(t+a)}_{\rm t}=\E^{-\I a\hat H}\ket{\psi(t)}_{\rm t}
\end{align}
where $\ket{\psi(t)}_{\rm t}$ represents a state at time $t$ and evolving in the $+t$ time direction.

The symmetry operations relevant to these translations are the parity inversion $\hat P$ and the time reversal $\hat T$ operations\footnote{We use the operator symbols $\hat P$ and $\hat T$ to represent the operations and the letters P and T to represent the corresponding symmetries.  Thus, if the system is invariant to the $\hat P$ operation it obeys the P symmetry.}  defined by Wigner \cite{Wigner}.  Parity inversion interchanges $x$ with $-x$, $y$ with $-y$ and $z$ with $-z$.  For example, $\hat P\ket{x}_{\rm x}=\ket{-x}_{\rm x}$ and $\hat T\ket{t}_{\rm t}=\ket{-t}_{\rm t}$.  The reverse of the translation in \eq{eq:x - translate by a} can be written as
\begin{align*}
        \ket{x-a}_{\rm x}&=\hat P\ket{-x+a}_{\rm x}=\hat P\E^{-\I a\hat p_x}\ket{-x}_{\rm x}=\hat P\E^{-\I a\hat p_x}\hat P^{-1}\ket{x}_{\rm x}\ .
\end{align*}
As $\hat P\hat p_x\hat P^{-1}=-\hat p_x$ we get
\begin{align*}
        \ket{x-a}_{\rm x}=\E^{\I a\hat p_x}\ket{x}_{\rm x}
\end{align*}
as expected directly from \eq{eq:x - translate by a}.  This shows that the generator of translations in either direction of the $x$ axis is the same.  The reverse of the translation in \eq{eq:t - translate by a} is somewhat different, however.  Consider
\begin{align}
        \ket{\phi(t-a)}_{\rm t}&=\hat T\ket{\phi(-t+a)}_{\rm t}
        =\hat T\E^{-\I a\hat H}\ket{\phi(-t)}_{\rm t}
        =\hat T\E^{-\I a\hat H}\hat T^{-1}\ket{\phi(t)}_{\rm t}\nonumber\\
        &=\E^{\I a\hat T\hat H\hat T^{-1}}\ket{\phi(t)}_{\rm t} \label{eq:t - translate by -a}
\end{align}
where $\ket{\phi(t)}_{\rm t}$ represents a state that evolves in the $-t$ direction and we have made use of the antiunitary nature of the time reversal operator, i.e. $\hat T\I\hat T^{-1}=-\I$, in the last line \cite{Wigner}.  In general $\hat T\hat H\hat T^{-1}\ne\hat H$ and so we set, for convenience,
\begin{align*}
        \hat H_B&=\hat T\hat H\hat T^{-1}\\
        \hat H_F&=\hat H
\end{align*}
where the subscripts $F$ and $B$ refer to the ``forwards'' and ``backwards'' direction of time corresponding to the $+t$ and $-t$ time directions, respectively.  If T symmetry is obeyed then
\begin{align*}
        \hat H_B=\hat H_F=\hat H\ ,   \qquad\mbox{(T symmetry)}
\end{align*}
and so there is a unique version of the Hamiltonian, whereas for T violation there is a different version of the Hamiltonian for each direction of time,
\begin{align*}
        \hat H_B\ne\hat H_F\ .\qquad\mbox{(T violation)}
\end{align*}
In general, we write \eq{eq:t - translate by a} and \eq{eq:t - translate by -a} as
\begin{align*}
        \ket{\psi(t+a)}_{\rm t}&=\E^{-\I a\hat H_F}\ket{\psi(t)}_{\rm t}\\
        \ket{\phi(t-a)}_{\rm t}&=\E^{\I a\hat H_B}\ket{\phi(t)}_{\rm t}\ . 
\end{align*}

The key point to be made here is that the generator of translations in space,  $\hat p_x$, is invariant (up to a sign change) under any of the C, P and T operations. In contrast, the generator of translations in time,  $\hat H$, is not invariant to the C, P and T operations, in general.  This underlies the statement in the Introduction that the symmetry violations occur over translations in time and not translations in space.

In the case of T violation we need to take care with using the correct Hamiltonian associated with each direction of time.  In particular, we need to apply the following principle:
\begin{principle} \label{p:time evolution}
Physical time evolution is represented by the operators $\E^{-\I a\hat H_F}$ and $\E^{\I a\hat H_B}$ for the forward ($+t$) and backward ($-t$) directions of time, respectively. The operations $\E^{\I a\hat H_F}$ and $\E^{-\I a\hat H_B}$ represent the mathematical inverse operation of  ``unwinding'' or ``backtracking'' the evolution produced by $\E^{-\I a\hat H_F}$ and $\E^{\I a\hat H_B}$, respectively.
\end{principle}
For example, $\E^{\I a\hat H_F}\ket{\psi(t+a)}_{\rm t}=\ket{\psi(t)}_{\rm t}$ represents unwinding the time evolution $\E^{-\I a\hat H_F}\ket{\psi(t)}_{\rm t}=\ket{\psi(t+a)}_{\rm t}$
whereas $\E^{\I a\hat H_B}\ket{\psi(t+a)}_{\rm t}$, which is not equal to  $\ket{\psi(t)}_{\rm t}$ in general, represents time evolution of $\ket{\psi(t+a)}_{\rm t}$ in the $-t$ direction. More details are given in Ref. \cite{PRSA}.

\section*{Appendix 3}
\addcontentsline{toc}{section}{Appendix 3}

In this Appendix we briefly discuss the mathematical construction of quantum virtual paths for the generic coordinate $w$.  Let the generators of translations be given by $\hat W_F$ and $\hat W_B$ for the $+w$ and $-w$ directions, respectively.
A quantum virtual path of the kind we want is given by \cite{PRSA}
\begin{align}  \label{eq:qvp}
   \ket{g}_N  \propto
     \frac{1}{2^N}\left(\E^{\I\hat W_B\delta w}\!+\!\E^{-\I\hat W_F\delta w}\right)^N\ket{0}_{\rm w}
\end{align}
where $\delta w$ is given by \eq{eq:delta w} and represents an increment in $w$ and $\ket{w}_{\rm w}$ represents a state for which $w$ is well-defined.\footnote{If $w$ represents a spatial coordinate then $\ket{w}_{\rm w}$ would be a corresponding spatial eigenstate.  For the case where $w$ represents the time coordinate, however, we only need $\ket{w}_{\rm w}$ to represent a well-defined time.  More details can be found in Ref.~\cite{PRSA}. } Expanding the power on the right side gives $2^N$ terms each with $N$ factors.  Each term represents a path comprising $N$ steps of $\delta w$ over the $w$ coordinate.  For example, a term of the form
\begin{align*}
        \ldots\E^{\I\hat W_B\delta w}\E^{-\I\hat W_F\delta w} \E^{\I\hat W_B\delta w}\E^{\I\hat W_B\delta w}\E^{-\I\hat W_F\delta w}\ket{0}_{\rm w}
\end{align*}
represents the object starting at the origin $w=0$ and then undergoing virtual displacements to $w=\delta w$, $w=0$, $w=-\delta w$, $w=0$ $w=-\delta w$ and so on.

It is relatively straightforward to show that the state $ \ket{g}_N$ in \eq{eq:qvp} approaches a Gaussian state in the limit of large $N$ when the discrete symmetry holds. To see this set $\hat W_B=\hat W_F=\hat W$ and use
\begin{align*}
     \exp(-A^2/2)=\lim_{N\to\infty}\cos^N(A/\sqrt{N})\
\end{align*}
to find
\begin{align*}
      \lim_{N\to\infty}\ket{g}_N  \propto \E^{-\hat W^2\sigma^2_{\rm w}}\ket{0}_{\rm w}\ ,
\end{align*}
and then, assuming that $\hat W$ has a complete orthonormal basis, rewrite this as the Fourier integral
\begin{align*}
      \lim_{N\to\infty}\ket{g}_N & \propto\int  \D w\ \E^{-w^2/4\sigma^2_{\rm w}} \E^{-\I \hat{W}w} \ket{0}_{\rm w}\\
        &= \int g(w)\ket{w}_{\rm w} \D w
\end{align*}
where $g(w)$ is given by
\begin{align*}
       g (w) =  \E^{-w^2/4\sigma^2_{\rm w}}\ .
\end{align*}
The square of this, $g^2(w)$,  is proportional to the bell-shaped probability distribution $P(w)$ represented in Fig.~\ref{fig:gaussian}.

\end{document}